\documentclass[12pt]{article}
\usepackage{amssymb,amsmath}

\newtheorem{theorem}{Theorem}[section]
\newtheorem{lemma}[theorem]{Lemma}
\newtheorem{proposition}[theorem]{Proposition}
\newtheorem{corollary}[theorem]{Corollary}
\newenvironment{proof}[1][Proof]{\begin{trivlist}
\item[\hskip \labelsep {\bfseries
#1}]}{\end{trivlist}}
\newenvironment{definition}[1][Definition]{\begin{trivlist}
\item[\hskip \labelsep {\bfseries
#1}]}{\end{trivlist}}
\newenvironment{example}[1][Example]{\begin{trivlist} \item[\hskip
\labelsep {\bfseries
#1}]}{\end{trivlist}}
\newenvironment{remark}[1][Remark]{\begin{trivlist}
\item[\hskip
\labelsep {\bfseries #1}]}{\end{trivlist}} \newcommand{\qed}{\nobreak
\ifvmode \relax \else
\ifdim\lastskip<1.5em \hskip-\lastskip
\hskip1.5em plus0em minus0.5em \fi \nobreak \vrule height0.75em width0.5em
depth0.25em\fi}
\newcommand{\E}{\textup{\bf E}}

\begin{document}

\begin{centering}

{\huge Canonical Transformations of Local}

{\huge \vspace{.5cm} Functionals and sh-Lie
Structures}
\\

\renewcommand{\thefootnote}{\fnsymbol{footnote}}
\vspace{1.4cm}
{\large Samer Al-Ashhab and Ron Fulp}\\

\vspace{.5cm}

 Department of Mathematics, North Carolina State University, Raleigh,
NC 27695-8205.\\
\it E-mail: ssalash@unity.ncsu.edu\\
E-mail: fulp@math.ncsu.edu\\
\rm
\vspace{.5cm}

\rm
\vspace{.5cm}
\vspace{.5cm}

\begin{abstract} In many Lagrangian field theories, there is a Poisson bracket
on the space of local functionals. One may identify the fields of such theories
as sections of a vector bundle. It is known that the Poisson bracket induces an
sh-Lie structure on the graded space of horizontal forms on the jet bundle of
the relevant vector bundle. We consider those automorphisms of the vector
bundle which induce mappings on the space of functionals preserving the Poisson
bracket and refer to such automorphisms as canonical automorphisms.

We determine how such automorphisms relate to the corresponding sh-Lie
structure. If a Lie group acts on the bundle via canonical automorphisms, there
are induced actions on the space of local functionals and consequently on the
corresponding sh-Lie algebra. We determine conditions under which the sh-Lie
structure induces an sh-Lie structure on a corresponding reduced space where
the reduction is determined by the action of the group. These results are not
directly a consequence of the corresponding theorems on Poisson manifolds as
none of the algebraic structures are Poisson algebras.

\end{abstract}

\end{centering}


\newpage

\section{Introduction} The dynamical ``equations of motion" of a Lagrangian
field theory are usually derived from a variational principle of ``least
action". Given a Lagrangian $L$, the action of $L$ is the functional $S$
defined by $$S(\phi)=\int_M L((j^n\phi)(x)) Vol_M$$ where $M$ is a manifold,
$\phi$ may be either a vector-valued function or a section of a vector bundle
$E$ over $M$, and $L$ is a real-valued function on some finite jet bundle
$J^nE.$ More generally, if $\pi:E\rightarrow M$ is a vector bundle and
$\pi^{\infty}:J^{\infty}E\rightarrow M$ is the corresponding prolongation of
$E,$ then a smooth function $P:J^{\infty}E \rightarrow {\bf R}$ is called a
{\bf local function} on $E$ provided that for some positive integer $n$ there
is a smooth function $P_n:J^nE\rightarrow {\bf R}$ such that $P=P_n\circ \pi_n$
where $\pi_n$ is the projection of $J^{\infty}E$ onto $J^nE.$ Thus all
Lagrangians are local functions on an appropriate bundle. To say that ${\cal
P}$ is a {\bf local functional} on $E$ means that ${\cal P}$ is a mapping from
a subspace of compactly supported sections of $E\rightarrow M$ into ${\bf R}$
such that $${\cal P}(\phi)=\int_M(P\circ j^{\infty}\phi)(x) Vol_M$$ for some
local function $P$ and for all such sections $\phi$ of $E.$

In imitation of Hamiltonian mechanics one postulates the existence of a
``Poisson bracket" on the space ${\cal F}$ of local functionals and then uses
it to develop a Hamiltonian theory of fields. This bracket is assumed to
satisfy the Jacobi identity and so defines a Lie algebra structure on the space
${\cal F}$. On the other hand there is no obvious commutative multiplication of
such functionals and consequently ${\cal F}$ is not a Poisson algebra. This is
such a well-known development that we may refer to standard monographs on the
subject. In particular we call attention to \cite{D91} and \cite{O86} for
classical expositions and to \cite{HT92} for a quantum field theoretic
development.

It was shown in \cite{BFLS98} that a Poisson bracket on the space of local
functionals induces what is known as sh-Lie structure on a part of the
bivariational complex which we refer to as the ``de Rham complex" on
$J^{\infty}E.$ This sh-Lie structure is given by three mappings $l_1,l_2,l_3$
defined on this complex. The mapping $l_2$ is skew-symmetric and bilinear, and
it may be regarded as defining a ``bracket" but one which generally fails to
satisfy the Jacobi identity. In fact $l_2$ satisfies the Jacobi identity if
$l_3=0.$ In a sense this sh-Lie structure is an anti-derivative of the Poisson
bracket.

In the present paper we intend to develop ideas related to canonical
transformations of these structures. Recall that if $M$ is a Poisson manifold
one says that a mapping  from $M$ to itself is a canonical transformation if
and only if it preserves the Poisson bracket defined on $C^{\infty}(M)$
\cite{MR94}. Moreover if one has a Lie group $G$ which acts on $M$ via
canonical transformations, one obtains a reduction of the brackets to a reduced
space $M/G$ in the presence of appropriate hypothesis. The space of local
functionals is not a Poisson algebra and so there is no underlying Poisson
manifold. The bracket $l_2$ is defined on the space of ``top" forms of the
``de Rham complex" which can be identified with the space of local functions on
$J^{\infty}E.$ This space is a commutative algebra under pointwise
multiplication, but $l_2$ does not satisfy the Jacobi identity and so again one
does not have a Poisson manifold.

We say that an automorphism of the bundle $E$ is {\em canonical} provided that
its induced mapping on the space of local functionals, ${\cal F}$, preserves
the Poisson bracket. We then determine how such automorphisms relate to the
sh-Lie structure on $J^{\infty}E.$ Finally we determine conditions under which
there exists an sh-Lie structure on a reduced graded space in the presence of a
Lie group which acts on $E$ via canonical automorphisms. \\

We apply these ideas to a Poisson sigma model. Poisson sigma models have proven
to be of interest in many areas of physics. In particular they have been used
to describe certain two-dimensional theories of gravity by Ikeda \cite{I94},
topological field theories by Schaller and Strobl \cite{SS94}, and to obtain a
path integral proof of Kontsevich's theorem on deformation quantization by
Catanneo and Felder \cite{CF99}. These are but a sample of the many authors who
have made important contributions relating to these model theories. \\

After presenting some background material in section 2, we find conditions for
the {\em induced} automorphisms on the space of local functionals to be {\em
canonical}. In section 3 we show how these induced canonical transformations
relate to the sh-Lie structure maps. In section 4 we assume that one has a Lie
group acting by canonical transformations on the space of functionals. We then
determine how this action relates to the sh-Lie structure and we find
conditions for the existence of an (induced) sh-Lie structure on a
corresponding reduced space. There is a brief discussion of functional
invariance in section 5. In section 6 we show how our formalism applies to a
specific Poisson sigma model due initially to Ikeda.

Clearly the questions dealt with in this paper relate more to the mathematical
structures induced by a Poisson bracket on the space of functionals rather than
to specific methods of solving dynamical field equations. Moreover we have
restricted our attention to a class of theories in which the Poisson bracket is
induced by a tensor $\omega$ which is scalar-valued rather than
differential-operator valued. Once we understand this restricted case more
fully we hope to extend these results to a larger class of theories for which
$\omega$ is differential operator valued.

Eventually we also intend to expand our scope to include fermionic theories
such as those in \cite{HT92}. Indeed, the sh-Lie formalism is particularly
well-suited to interact with super-field theories such as those needed to
describe the Batalin-Vilkovisky approach to BRST cohomology. Once anti-fields
are introduced, our vector bundle $E$ can be modified in such a manner that
both the bosonic fields studied here and the fermionic (anti-)fields become
sections of the new bundle. In this context the Batalin-Vilkovisky anti-bracket
is none other than our Poisson bracket of local functionals with appropriate
grading. Thus we expect the modifications of this work to the latter case to be
minimal. In fact this work is motivated by both the classical field theories
such as those described in \cite{O86} and the super-fields developed in
\cite{HT92}. This approach has proven its worth in investigations such as those
found in \cite{BH96} and \cite{BH93}.

\section{Bundle automorphisms preserving the Poisson structure on the space of
functionals} \label{BAS} \subsection{Background material} In this section we
introduce some of the terminology and concepts that are used in this work, in
addition to some of the simpler results that will be needed. Our exposition and
notation closely follows that in \cite{BFLS98}. First let $E \to M$ be a vector
bundle where the base space $M$ is an $n$-dimensional manifold and let
$J^\infty E$ be the infinite jet bundle of $E.$ The restriction of the infinite
jet bundle over  an appropriate open set $U\subset M$ is trivial with fiber an
infinite dimensional vector space $V^\infty$.  The bundle \begin{eqnarray*}
\pi^\infty : J^\infty E_U=U\times V^\infty \rightarrow U \end{eqnarray*} then
has induced coordinates given by \begin{eqnarray*}
(x^i,u^a,u^a_i,u^a_{i_1i_2},\dots,). \end{eqnarray*} We use multi-index notation
and the summation convention throughout the paper. If $j^{\infty}\phi$ is the
section of $J^{\infty}E$ induced by a section $\phi$ of the bundle $E$, then
$u^a\circ j^{\infty}\phi=u^a\circ \phi$ and $$u^a_I\circ j^{\infty}\phi=
(\partial_{i_1}\partial_{i_2}...\partial_{i_r})(u^a\circ j^{\infty}\phi)$$
where $r$ is the order of the symmetric multi-index
$I=\{i_1,i_2,...,i_r\}$, with the convention that, for $r=0$, there are no
derivatives. For more details see \cite{A96} and \cite{KV98}.

Let $Loc_E$ denote the algebra of local functions where a local function on
$J^\infty E$ is defined to be the pull-back of a smooth function on some finite
jet bundle $J^p E$ via the projection from $J^\infty E$ to $J^p E$. Let
$Loc_E^0$ denote the subalgebra of $Loc_E$ such that $P \in Loc_E^0$ iff
$(j^\infty \phi)^* P$ has compact support for all $\phi \in \Gamma E$ with
compact support, where $\Gamma E$ denotes the set of sections of the bundle $E
\to M$. The de Rham complex of differential forms $\Omega^*(J^{\infty}E,d)$ on
$J^{\infty}E$ possesses a differential ideal, the ideal ${ C}$ of contact forms
$\theta$ which satisfy $(j^{\infty}\phi)^* \theta=0$ for all sections $\phi$
with compact support. This ideal is generated by the contact one-forms, which
in local coordinates assume the form $\theta^a_J=du^a_J-u^a_{iJ}dx^i$.

Using the contact forms, we see that the complex $\Omega^*(J^{\infty}E,d)$
splits as a bicomplex $\Omega ^{r,s}(J^ \infty E)$ (though the finite level
complexes $\Omega^*(J^pE)$ do not), where $\Omega ^{r,s}(J^ \infty E)$ denotes
the space of differential forms on $J^\infty E$ with $r$ horizontal components
and $s$ vertical components. The bigrading is described by writing a
differential $p$-form $\alpha=\alpha_{IA}^{\bf J}(\theta^A_{\bf J} \wedge
dx^I)$ as an element of $\Omega^{r,s}(J^{\infty}E)$, with $p=r+s$, and
\begin{eqnarray*} dx^I=dx^{i_1}\wedge...\wedge dx^{i_r} \quad \quad
  \quad \theta^A_{\bf J}=\theta^{a_1}_{J_1}\wedge...\wedge
  \theta^{a_s}_{J_s}.
\end{eqnarray*}

Now let $C_0$ denote the set of contact one-forms of {\em order zero}. Contact
one-forms of order zero satisfy $(j^{1}\phi)^*(\theta)=0$ and, in local
coordinates, they assume the form $\theta^a= du^a-u^a_idx^i$. Notice that both
$C_0$ and $\Omega ^{n,1} = \Omega ^{n,1}(J^ \infty E)$ are modules over
$Loc_E$. Let $\Omega ^{n,1}_0$ denote the subspace of $\Omega ^{n,1}$ which is
locally generated by the forms $\{(\theta ^a \wedge d^nx)\}$ over $Loc_E$. We
assume the existence of a mapping, $\omega$, from $\Omega ^{n,1}_0 \times
\Omega ^{n,1}_0$ to $Loc_E$, such that $\omega$ is a skew-symmetric module
homomorphism in each variable separately. In local coordinates let $\omega
^{ab} = \omega (\theta ^a \wedge \nu,\theta ^b \wedge \nu)$, where $\nu$ is a
volume element on $M$ (notice that in local coordinates $\nu$ takes the form
$\nu = f d^nx = f dx^1 \wedge dx^2 \wedge ... \wedge dx^n$ for some function
$f: U \to {\mathbf R}$ and $U$ is a subset of $M$ on which the $x^i$'s are
defined). \bigskip

{\it We will assume throughout this paper that $\omega$ satisfies the
conditions that make our Poisson bracket, which will be defined soon, satisfy
the Jacobi identity.}

Define the operator $D_i$ (total derivative) by $\displaystyle D_i =
\frac{\partial}{\partial x^i} + u^a_{iJ}\frac{\partial}
{\partial u^a_J}$ (recall we assume the summation convention, i.e.,
the sum is over all $a$ and multi-index $J$), and recall that the
Euler-Lagrange operator
maps
$\Omega ^{n,0}(J^\infty E)$ into $\Omega ^{n,1}(J^\infty E)$ and is
defined
by $$\E (P \nu)=\E _a(P)(\theta ^a \wedge \nu)$$ where $P \in Loc_E, \nu$
is a
volume
form on the base manifold $M$,
and the
components $\E_a(P)$ are given by $$\E_a(P)=(-D)_I(\frac{\partial P}
{\partial u^a_I}).$$ For simplicity of notation we may use $\E(P)$ for $\E
(P\nu)$.
We will also use $\tilde{D}_i$
for $\displaystyle \frac{\partial}{\partial \tilde{x}^i} +
\tilde{u}^a_{iJ}\frac{\partial}{\partial \tilde{u}^a_J}$
and $\tilde{\E}_a(P)$ for
$\displaystyle (-\tilde{D})_I(\frac {\partial P} {\partial
\tilde{u}^a_I})$
so that $\E(P)
= \tilde{\E}_a(P) (\tilde{\theta} ^a \wedge \nu)$ in the
$(\tilde{x}^{\mu},\tilde{u}^a)$ coordinate system. \\

Let $\Omega ^{k,l}_c(J^ \infty E)$ be the subspace of $\Omega ^{k,l}(J ^\infty
E)$, for $\{k,l\} \neq \{n,0\}$, such that $\alpha \in \Omega ^{k,l}_c(J^
\infty E)$ iff $(j^\infty \phi)^* \alpha$ has compact support for all $\phi \in
\Gamma E$ with compact support, and let $\Omega ^{n,0}_{c}(J^ \infty E)$ be the
subspace of $\Omega ^{n,0}(J^ \infty E)$ such that $P\nu \in \Omega ^{n,0}_{c}
(J^ \infty E)$ iff $(j^\infty \phi)^*(P\nu)$ and $(j^\infty \phi)^* \E_a(P)$
have compact support for all $\phi \in \Gamma E$ with compact support and for
all $a$. We are interested in the complex $$ 0 \to \Omega ^{0,0}_c(J^\infty E)
\to \Omega ^{1,0}_c(J^\infty E) \to \cdots \to \Omega ^{n-1,0}_c(J^\infty E)
\to \Omega ^{n,0}_{c}(J^\infty E)$$ with the differential $d_H$ defined by $d_H
= dx^i D_i$, i.e., if $\alpha = \alpha _I dx^I$ then $d_H \alpha = D_i \alpha
_I dx^i \wedge dx^I$. Notice that this complex is exact whenever the base
manifold $M$ is contractible (e.g. see \cite{BT82}). \\

Now let $\mathcal F$ be the space of functionals where $\mathcal P \in
\mathcal
F$ iff $\displaystyle \mathcal P=\int_M P \nu$ for some $P \in Loc_E^0$,
and
define a Poisson bracket on $\mathcal F$ by $$\{\mathcal P,\mathcal
Q\}(\phi)=
\int_M [\omega (\E(P),\E(Q))\circ j\phi] \nu,$$
where $\phi \in \Gamma E$, $\nu$ is a volume form on $\displaystyle M, \mathcal
P=
\int_M P \nu, \mathcal Q=\int_M Q \nu,$ and $P, Q \in Loc_E^0$. Using
local
coordinates $(x^\mu,u^a_I)$ on $J^\infty
E$, observe that for $\phi \in \Gamma E$ such that the support of $\phi$
lies
in the domain $\Omega$ of some chart
$x$ of $M$, one has
$$\{\mathcal P,\mathcal Q\}(\phi) =
\int_{x(\Omega)} ([\omega ^{ab} \E_a(P)\E_b(Q)] \circ j\phi \circ x^{-1})
(x^{-1})^*(\nu)$$
where $x^{-1}$ is the inverse of $x=(x^\mu)$. \\

{\em We assume that $\omega$ satisfies the necessary conditions for the above
bracket to satisfy the Jacobi identity, e.g. see \cite{O86}. Notice that it
follows from the identity (7.11) of \cite{O86} that  the bracket satisfies the
Jacobi identity if the  skew-symmetric matrix $\{\omega^{ab}\}$ is a Poisson
tensor in the sense that:
$$\omega^{cd}\frac {\partial \omega^{ab}}{\partial u^d}+
\omega^{ad}\frac {\partial \omega^{bc}}{\partial u^d}+
\omega^{bd}\frac {\partial \omega^{ca}}{\partial u^d}=0,$$
where $\{u^a\}$ are coordinates on the fiber of the trivial bundle $E.$ This
condition is met in the case of the Poisson sigma model, which we include later
in the paper, and more generally for any trivial vector bundle with a Poisson
structure on its fibers.} \\

The functions $P$ and $Q$ in our definition of the Poisson bracket of local
functionals are representatives of $\mathcal P$ and $\mathcal Q$ respectively,
since generally these are not unique. In fact $\mathcal F \simeq H^n_c(J^\infty
E)$, where $H^n_c(J^\infty E) = \Omega ^{n,0}_{c}(J^\infty E)/(\textup{im}
d_H\bigcap\Omega ^{n,0}_{c}(J^\infty E))$ and im$d_H$ is the image of the
differential $d_H$. \\

Let $\psi:E\rightarrow E$ be an automorphism, sending fibers to fibers, and let
$\psi _M:M\rightarrow M$ be the induced diffeomorphism of $M$. Notice that
$\psi$ induces an automorphism $j\psi: J^{\infty}E \rightarrow J^{\infty}E$
where $$(j\psi)((j^\infty \phi)(p))=j(\psi \circ \phi \circ \psi ^{-1}_M)(\psi
_M(p)),$$ for all $\phi \in \Gamma E$ and all $p$ in the domain of $\phi$.  In
these coordinates the independent variables transform via $\tilde{x}^\mu = \psi
_M^\mu(x^\nu)$. Local coordinate representatives of $\psi _M$ and $j\psi$ may
be described in terms of charts $(\Omega,x)$ and $(\tilde{\Omega},\tilde{x})$
of $M$, and induced charts $((\pi ^\infty)^{-1} (\Omega),(x^{\mu},u^a_I))$ and
$((\pi ^\infty)^{-1} (\tilde{\Omega}),(\tilde{x}^{\mu},\tilde{u}^a_I))$ of $J^
\infty E$.

\begin{remark} In section 4 we will consider (left) Lie group actions on $E$
and their induced (left) actions on $J^ \infty E.$ Such actions are defined by
homomorphisms from the group into the group of automorphisms of $E.$
\end{remark}

\begin{definition} $\omega:\Omega ^{n,1}_0 \times \Omega ^{n,1}_0 \rightarrow
Loc_E$ is {\em covariant} with respect to an automorphism $\psi:E \rightarrow
E$ of the above form iff $$\omega((j\psi)^* \theta,(j\psi)^* \theta ') =
(\textup{det}\psi _M)(j\psi)^*(\omega(\theta, \theta ')),$$ for all
$\theta,\theta ' \in \Omega ^{n,1}_0(J^ \infty E)$. \end{definition}

\noindent
Observe that
\begin{eqnarray*}
(j\psi)^*\tilde{\theta} ^a
& = &
(j\psi)^*(d\tilde{u}^a-\tilde{u}^a_\mu d\tilde{x}^\mu) \\ & = &
d(\tilde{u}^a\circ j\psi) - (\tilde{u}^a_\mu\circ j\psi)
d(\tilde{x}^\mu\circ
j\psi) \\
& = &
\frac{\partial \psi_E^a}{\partial x^\nu} dx^\nu + \frac{\partial
\psi_E^a}{\partial u^b} du^b - (\frac{\partial \psi_E^a}{\partial
x^\nu}+\frac{\partial \psi_E^a}{\partial u^b}
u_\nu^b)(J^{-1})_\mu ^\nu \frac{\partial \tilde{x}^\mu \circ j\psi}
{\partial x^\lambda} dx^ \lambda \\
& = &
\frac{\partial \psi_E^a}{\partial x^\nu} dx^\nu + \frac{\partial
\psi_E^a}{\partial u^b} du^b - \frac{\partial \psi_E^a}{\partial x^\nu}dx^
\nu - \frac{\partial \psi_E^a}{\partial u^b} u_\nu^b dx^\nu \\ & = &
\frac{\partial \psi_E^a}{\partial u^b}(du^b - u_\nu^b dx^\nu) \\ & = &
\frac{\partial \psi_E^a}{\partial u^b} \theta ^b \end{eqnarray*}
where we have assumed that $\psi_E^a = \tilde{u}^a \circ \psi$ and $J$ is the
Jacobian matrix of the transformation $\tilde{x}^\nu = \psi _M^\nu
(x^\mu)$.

\begin{lemma} \label{cov}
The following are equivalent \\
(i) $\omega((j\psi)^* \theta,(j\psi)^* \theta ') = (\textup{det}\psi _M)
(j\psi)^* (\omega(\theta,\theta ')),$ for all $\theta,\theta ' \in \Omega
^{n,1}_0(J^\infty E)$.
\\
(ii) $\displaystyle \tilde{\omega}^{ab}\circ j\psi = (\textup{det}\psi _M)
\omega ^{cd}\frac{\partial \psi_E^a}{\partial u^c} \frac{\partial
\psi_E^b}{\partial u^d}.$ \end{lemma}

\begin{proof}
Notice that
\begin{eqnarray*}
\textup{det}\psi _M(j\psi)^* (\omega(\E(P),\E(Q)) & = &
(\textup{det}\psi _M)\omega(\E(P),\E(Q))\circ j\psi \\ & = &
\textup{det}\psi _M[\omega(\tilde{\theta} ^a \wedge \nu,\tilde {\theta} ^b
\wedge \nu) \tilde{\E}_a(P) \tilde{\E}_b(Q)] \circ j\psi \\ & = &
\textup{det}\psi _M(\tilde{\omega} ^{ab} \circ j\psi)(\tilde{\E}_a(P)
\circ
j\psi)(\tilde{\E}_b(Q) \circ j\psi)
\end{eqnarray*}
and that
\begin{eqnarray*}
(j\psi)^*(\E(P))
& = &
(j\psi)^* (\tilde{\E}_a(P)(\tilde{\theta}^a \wedge \nu)) \\ & = &
(\tilde{\E}_a(P)\circ j\psi) \frac{\partial \psi_E^a}{\partial u^c}
(\textup{det}\psi _M) (\theta ^c \wedge \nu). \end{eqnarray*}
Now
\begin{eqnarray*}
\omega((j\psi)^*\E(P),(j\psi)^*\E(Q))
& = &
(\textup{det}\psi _M)^2 \frac{\partial \psi^a_E}{\partial u^c}
\frac{\partial \psi^b_E}{\partial u^d} (\tilde{\E}_a(P) \circ j\psi)
(\tilde{\E}_b(Q) \circ j\psi) \\
& &
\omega(\theta ^c \wedge \nu,\theta ^d \wedge \nu) \\ & = &
(\textup{det}\psi _M)^2 \frac{\partial \psi^a_E}{\partial u^c}
\frac{\partial \psi^b_E}{\partial u^d} \omega ^{cd}(\tilde{\E}_a(P) \circ
j\psi)
(\tilde{\E}_b(Q) \circ j\psi).
\end{eqnarray*}
\noindent
Hence $\omega((j\psi)^*\E(P),(j\psi)^*\E(Q))=
(\textup{det}\psi _M)(j\psi)^* (\omega(\E(P),\E(Q))$ for all $P, Q$ in
$Loc_E$
iff $\displaystyle (\tilde{\omega} ^{ab} \circ j\psi) = (\textup{det}\psi
_M)
\frac{\partial \psi^a_E}{\partial u^c}
\frac{\partial \psi^b_E}{\partial u^d} \omega ^{cd}$. \qed \end{proof}

\subsection{Automorphisms preserving the Poisson structure}

Let $L:J^{\infty}E \rightarrow {\bf R}$ be a Lagrangian in $Loc_E$ (generally
we will assume that any element of $Loc_E$ is a Lagrangian). Let $\hat{L}
= L\circ (x^{\mu},u^a_I)^{-1}$ and let $\tilde{L} = L\circ
(\tilde{x}^{\mu},\tilde{u}^a_I)^{-1}$. Then, in local coordinates, $\tilde{L}$
is related to $\hat{L}$
by the equation $$(\tilde{L} \circ j\bar{\psi}) \textup{det}(J) = \hat{L},$$
where $j\bar{\psi} = (\tilde{x}^{\nu},\tilde{u}^b_K) \circ
(x^{\mu}, u^a_I)^{-1}$ and $J$ is the Jacobian matrix of the transformation
$\psi _M= \tilde{x}^\nu \circ (x^\mu)^{-1}$.
With abuse of notation we may assume coordinates and charts are the same and
write $\tilde{x}^\nu = \psi _M(x^\mu)$.
For simplicity, we have also assumed that
$\psi _M$ is orientation-preserving. In this case the functional
$$\tilde{\mathcal L}=\int_{\tilde{\Omega}} \tilde{L} d^n\tilde{x}$$ is the
transformed form of the functional $$\hat{{\mathcal L}}=\int_{\Omega} \hat{L}
d^nx$$ where $\hat{L}$ and $\tilde{L}$ are related as above, $\Omega$ is the
domain of integration and $\tilde{\Omega}$ is the transformed domain under
$j\bar{\psi}$ (see \cite{O86} pp.249-250). Notice that both of these are local
coordinate expressions of the equation $\displaystyle {\mathcal L} = \int_M L
\nu$, for appropriately restricted charts. Now suppose that $\psi$ is an
automorphism of $E$, $j\psi$ its induced automorphism on $J^\infty
E$, and $\psi _M$ its induced (orientation-preserving) diffeomorphism on 
$M$. Also suppose that
$\hat{L}$ and $\tilde{L}$ are two Lagrangians related by the equation
$(\tilde{L} \circ j\psi)\textup{det}(\psi _M) = \hat{L}$. We have:

\begin{lemma} \label{Lagrangian} Let $P$ be a Lagrangian as above, then
\begin{equation}
\label{ELI}
\E_a((P\circ j\psi)\textup{det}(\psi _M)) =\textup{det}(\psi _M) \frac{\partial
\psi^c_E}
{\partial u^a} (\tilde{\E}_c(P)\circ j\psi). \end{equation}
\end{lemma}

\begin{proof} First notice that
$\displaystyle \E_{u^a}(\hat{L}) = \textup{det}(\psi _M)
\frac{\partial \psi^c_E} {\partial u^a} (\E_{\tilde{u}^c}
(\tilde{L}) \circ j\psi)$ (see \cite{O86} pp.250).
But $(\tilde{L} \circ j\psi)\textup{det}(\psi _M) = \hat{L}$. The
identity $\ref {ELI}$
follows by letting $P=\tilde{L}$. {\em Notice that this is justified
since $\tilde{L}$ is arbitrary in the sense that given any $L'$ there
exists
an $\hat{L}$ derived from a Lagrangian
$L$ as above such that $(L' \circ j\psi)
\textup{det}(\psi _M) = \hat{L}$ since $j\psi$ is an automorphism}. \qed
\end{proof}

Let $\hat{\psi}$ denote the mapping representing the induced action of the
automorphism on sections of $E$, i.e., $\hat{\psi}: \Gamma E \rightarrow \Gamma
E$ where $\hat{\psi}(\phi) = \psi \circ \phi \circ \psi ^{-1}_M$ and $\phi$ is
a section of $E$. This induces a mapping on the space of local functionals
given by

\begin{eqnarray*}
(\mathcal P \circ \hat{\psi})(\phi)
& = &
\mathcal P (\psi \circ \phi \circ \psi ^{-1}_M) \\ & = &
\int_M [P \circ j(\psi \circ \phi \circ \psi ^{-1}_M)] \nu \\ & = &
\int_M [P \circ j\psi \circ j\phi \circ \psi ^{-1}_M)] \nu \\ & = &
\int_M [P \circ j\psi \circ j\phi] (\textup{det} \psi _M) \nu, \\
\end{eqnarray*}
where $$\mathcal P (\phi)=\int_M (P \circ j\phi) \nu,$$ and $\phi$ is a
section
of $E$. \\

{\em We find conditions on those automorphisms of the space of functionals
under which the Poisson structure is
preserved}. \\

\noindent Recall that
$\displaystyle \{\mathcal P,\mathcal Q\} = \int_M \omega(\E(P),\E(Q))
\nu,$
and hence
\begin{equation}
\{\mathcal P,\mathcal Q\}(\phi) = \int_M [\omega(\E(P),\E(Q))\circ j\phi]
\nu.
\end{equation}

\noindent
Now
\begin{equation}
\{\mathcal P\circ \hat{\psi},\mathcal Q \circ \hat{\psi} \}(\phi)=
(\{\mathcal P,\mathcal Q\} \circ \hat{\psi})(\phi) \end{equation}

\noindent
is equivalent to \\

$\displaystyle \int_M [\omega(\E((P \circ j\psi)\textup{det}\psi _M),\E((Q
\circ
j\psi) \textup{det}\psi _M)) \circ j\phi] \nu = $ \\ $$\int_M
[(\omega(\E(P),\E(Q))\circ j\psi \circ j\phi)\textup{det}\psi _M]\nu,$$

\noindent
but since this holds for all sections $\phi$ of $E$ it is equivalent to
$$\omega(\E((P \circ j\psi) \textup{det}\psi _M),\E((Q \circ j\psi)
\textup{det}\psi _M)) =
(\omega(\E(P),\E(Q))\circ j\psi) \textup{det}\psi _M$$
up to a {\em divergence}.
The last equation is equivalent to
$$\omega ^{ab}\E_b((Q\circ j\psi)\textup{det}\psi _M)\E_a((P\circ j\psi)
\textup{det}\psi _M)=([\tilde{\omega}^{ab}\tilde{\E}_b(Q) \tilde{\E}_a(P)]
\circ
j\psi) \textup{det}\psi _M$$
up to a divergence, or (Lemma $\ref{Lagrangian}$) \\

$\displaystyle \omega ^{ab}(\textup{det}\psi _M)^2 \frac{\partial \psi^d_E}
{\partial u^b} \frac{\partial \psi^c_E}{\partial u^a}(\tilde{\E}_d(Q) \circ
j\psi)
(\tilde{\E}_c(P) \circ j\psi) = $ \\
$$(\tilde{\omega}^{ab} \circ j\psi) (\tilde{\E}_b(Q) \circ j\psi)
(\tilde{\E}_a(P) \circ j\psi) (\textup{det}\psi _M)$$
up to a divergence. Finally, since the
last equation is true for all $P$ and $Q$ it is equivalent to
$$\displaystyle \tilde{\omega}^{ab}\circ j\psi = (\textup{det}\psi _M)
\omega ^{cd}\frac{\partial \psi_E^a}{\partial u^c}
\frac{\partial \psi_E^b}{\partial u^d}$$
which is equivalent to
the covariance of $\omega$. (Notice that if the last equality does not hold
then by some choice of $P$ and $Q$
the equations above will not hold up to a divergence.)
We have established the following:

\begin{theorem} \label {canonical}  Let $\psi:E \rightarrow E$ be an
automorphism of $E$ sending fibers to fibers, and let $\Psi: {\mathcal F}
\rightarrow {\mathcal F}$ be the induced mapping defined by $\Psi({\mathcal P})
= \mathcal P \circ \hat{\psi}$ (noting that $\mathcal P \circ \hat{\psi}$ is
defined as above) where $\hat{\psi}:\Gamma E \rightarrow \Gamma E$ is given by
$\hat{\psi}(\phi) = \psi \circ \phi \circ \psi _M^{-1}$. Then $\Psi$ is
canonical in the sense that $$\{\Psi(\mathcal P),\Psi(\mathcal
Q)\}=\Psi(\{\mathcal P,\mathcal Q\})$$ for all $\mathcal P, \mathcal Q \in
\mathcal F$ iff $\omega$ is covariant with respect to $\psi$. \end{theorem}

\begin{definition}
An automorphism $\psi$ of $E$ is {\em canonical} provided the induced mapping
$\Psi:{\cal F}\longrightarrow {\cal F}$ is canonical (in the sense of the
preceding Theorem).
\end{definition}

\begin{example}
Consider $M={\mathbf R}, E={\mathbf R} \times {\mathbf R}^2$, and let
\[ \omega = \left( \begin{array}{cc}
0 & 1 \\
-1 & 0 \end{array} \right). \]
Consider the action $\psi_g:E \to E$ defined by $\psi _g(x,u^1,u^2) = (x,
g\cdot
(u^1,u^2))$ for some $g \in SO(2)$. It can easily be shown that $\omega$
is
covariant with respect to $\psi_g$ and hence the induced action is
canonical. As
an illustration let $g$ be the counter-clockwise rotation by $90^o$ (so
that $g
u^1 = u^2$ and $gu^2=-u^1$), and let
$\displaystyle \mathcal P = \int_{\mathbf R} P(u^1) dx$ and $\displaystyle
\mathcal Q =
\int_{\mathbf R} Q(u^2) dx$ for
real-valued differentiable functions $P$ and $Q$. Then $\displaystyle \Psi
(\mathcal P) = \int_{\mathbf R} P(u^2) dx, \Psi(\mathcal Q) = \int_{\mathbf R}
Q(-u^1) dx$ and
$\displaystyle \{\Psi(\mathcal P),\Psi(\mathcal Q)\} = - \int_{\mathbf R}
P'(u^2)Q'(-u^1) dx$.
On the other hand
$\displaystyle \{\mathcal P,\mathcal Q\} = - \int_{\mathbf R}
P'(u^1)Q'(u^2) dx$, so that
$\displaystyle \Psi(\{\mathcal P,\mathcal Q\}) = - \int_{\mathbf R}
P'(u^2)Q'(-u^1) dx$.
\qed
\end{example}

\noindent The following will be needed in our subsequent work.
\begin{proposition} \label {commute}
$d_H((j\psi) ^* \eta)=(j\psi) ^*(d_H\eta)$ for $\eta \in \Omega ^{m,0}$,
$m$ arbitrary.
\end{proposition}
\begin{proof}
Let $\eta = \alpha _I d\tilde{x}^I$ where each $\alpha _I: J^\infty
E\rightarrow {\mathbf R}$.
We assume that the $\alpha _I$'s depend on the transformed variables
$(\tilde{x}^j,\tilde{u}^b_K)$. Then \\
\begin{eqnarray*}
d_H((j\psi)^*(\alpha))
& = &
d_H((\alpha _I \circ j\psi) d(\tilde{x}^I\circ\psi)) \\
& = &
\displaystyle \{D_i(\alpha _I \circ j\psi)\} dx^i \wedge
d(\tilde{x}^I\circ\psi) \\
& = &
\{ (\frac{\partial \alpha _I}{\partial \tilde{x}^j}\circ j\psi)
\frac{\partial
(\tilde{x}^j\circ\psi)}{\partial x^i} +
(\frac{\partial \alpha _I}{\partial \tilde{u}^b_K} \circ j\psi)
\frac{\partial j\psi ^b_K}{\partial x^i} + \\
& &
u^a_{iI} (\frac{\partial \alpha _I}{\partial \tilde{u}^b_K} \circ j\psi)
\frac{\partial j\psi ^b_K}{\partial u^a_I}\} dx^i \wedge
d(\tilde{x}^I\circ\psi) \\
& = &
\{ (\frac{\partial \alpha _I}{\partial \tilde{x}^j} \circ j\psi)
d(\tilde{x}^j\circ\psi) +
(\frac{\partial \alpha _I}{\partial \tilde{u}^b_K} \circ j\psi) \cdot (
\frac{\partial j\psi ^b_K}{\partial x^i} + u^a_{iI} \frac{\partial j\psi
^b_K}{\partial u^a_I}) \\
& &
dx^i \}\wedge d(\tilde{x}^I\circ\psi) \\
& = &
\{ (\frac{\partial \alpha _I}{\partial \tilde{x}^j} \circ j\psi) +
(\frac{\partial \alpha _I}{\partial \tilde{u}^b_K} \circ j\psi) \cdot (
\frac{\partial j\psi ^b_K}{\partial x^i} + u^a_{iI} \frac{\partial j\psi
^b_K}{\partial u^a_I} ) (J^{-1})^i_j\} \\
& &
d(\tilde{x}^j\circ\psi) \wedge d(\tilde{x}^I\circ\psi) \\
& = &
\{ (\frac{\partial \alpha _I}{\partial \tilde{x}^j} \circ j\psi) +
(\frac{\partial \alpha _I}{\partial \tilde{u}^b_K} \circ j\psi) \cdot (
\tilde{u}^b_{Kj} \circ j\psi )\} \\ & &
d(\tilde{x}^j\circ\psi) \wedge d(\tilde{x}^I\circ\psi) \\
& = &
\{(\tilde{D}_j \alpha _I) \circ j\psi\} (d(\tilde{x}^j\circ\psi) \wedge
d(\tilde{x}^I\circ\psi))
= (j\psi)^*(d_H(\alpha)),
\end{eqnarray*}

\noindent
where we assumed that $j\psi ^b_K = \tilde{u}^b_K \circ j\psi$ and $J$ is the
Jacobian matrix of the transformation $\psi _M^\nu (x^\mu)$ as before.
\qed
\end{proof}

\section{Canonical automorphisms and sh-Lie algebras} In this section we
consider the structure maps of the sh-Lie algebra on the horizontal complex
\{$\Omega ^{i,0}$\}. Throughout the remainder of the paper we assume the
horizontal complex is exact. A complete description of these maps can be found
in \cite{BFLS98} or one of the references therein; however it is useful to give
a brief overview.

\subsection{Overview of sh-Lie algebras} Let ${\cal F}$ be a vector space and
$(X_*,l_1)$ a homological resolution thereof, i.e., $X_*$ is a graded vector
space, $l_1$ is a differential and lowers the grading by one with ${\cal
F}\simeq H_0(l_1)$ and $H_k(l_1)=0$ for $k>0$. The complex $(X_*,l_1)$ is
called the resolution space.  (We are {\it not} using the term `resolution' in
a categorical sense.) Consider a homological resolution of the space ${\cal F}$
of local functionals as in \cite{BFLS98}. In the field theoretic framework
considered in \cite{BFLS98} it was shown that under certain hypothesis (see the
Theorem below) the Lie structure defined by the Poisson bracket on ${\cal F}$
induces an sh-Lie structure on the graded vector space
$X_i=\Omega^{n-i,0}(J^{\infty}E),$ for $0\leq i<n$ and
$X_n=\Omega^{0,0}(J^{\infty}E).$ For completeness we give the definition of
sh-Lie algebras and include a statement of the relevant Theorem.

\begin{definition} An sh Lie structure on a graded vector space $X_*$ is a
collection of linear, skew-symmetric maps $l_k: \bigotimes ^k X_* \to X_*$ of
degree $k-2$ that satisfy the relation $$\sum_{i+j=n+1}\sum_{unsh(i,n-i)}
e(\sigma)(-1)^ \sigma (-1)^{i(j-1)}l_j(l_i
(x_{\sigma(1)},\cdots,x_{\sigma(i)}),\cdots,x_{\sigma(n)}) = 0,$$ where $ 1
\leq i,j$. \end{definition} Notice that in this definition $e(\sigma)$ is the
Koszul sign which depends on the permutation $\sigma$ as well as on the degree
of the elements $x_1,x_2,\cdots,x_n$ (see for example \cite{LM95}).

\begin{remark} Although this may seem to generally be a rather complicated
structure, it simplifies drastically in the case of field theory where, aside
from the differential $l_1=d_H,$ the only non-zero maps are $l_2$ and $l_3$ in
degree 0. \end{remark}

The Theorem relevant to field theory depends on the existence of a linear
skew-symmetric map $\tilde{l}_2: X_0 \otimes X_0 \to X_0$ (in our case the
the Poisson bracket will be the integral of this mapping as we will see in
detail shortly) satisfying conditions $(i)$ and $(ii)$ below. These conditions
are all that is needed in order that an sh Lie structure exists.

\begin{theorem} \label{shLie}
A skew-symmetric linear map $\tilde{l}_2: X_0 \otimes X_0 \to X_0$ that
satisfies
conditions (i) and (ii) below extends to an sh Lie structure on the
resolution space $X_*$;\\
$(i) \quad \tilde{l}_2(c,b_1) = b_2$ \\
$\displaystyle (ii) \sum _{\sigma \in unsh(2,1)} (-1)^\sigma
\tilde{l}_2(\tilde{l}_2(c_{\sigma(1)},c_{\sigma(2)}),c_{\sigma(3)}) =b_3$
\\
where $c, c_1, c_2, c_3$ are cycles and $b_1, b_2, b_3$ are boundaries in
$X_0$.
\end{theorem}

{\em Subsequently we will suppress
some of the notation and assume
the summands are over the appropriate shuffles with their corresponding
signs}. \\

We also assume the existence of a chain homotopy $s$ which satisfies $-s \circ
l_1=1 + l_1 \circ s$. Using this chain homotopy one can define $l_3$ in degree 0
by the composition $s\circ\tilde l_2 \circ\tilde l_2$ which we may write simply
as $s\tilde l_2 \tilde l_2$ (we assume the sum over the three unshuffles for
$\tilde l_2 \tilde l_2$ with their corresponding signs).

\subsection{The effect of a canonical automorphism on the structure maps of the
sh-Lie algebra} To apply the Theorem in the last subsection, we need a
candidate for $\tilde l_2.$ We may define such a mapping on $\Omega ^{n,0}$ by
\begin{equation} \label{l2} \tilde{l}_2(P\nu ,Q\nu)= \omega ^{ab}\E_a(P)\E_b(Q)
\nu =\omega(\E(P),\E(Q)) \nu. \end{equation} Recall that for each automorphism
$\psi$ we have $(j\psi)^*(P\nu) = (P\circ j\psi) (\textup{det}\psi _M) \nu$.
Therefore \begin{equation} \label{l2P}
\tilde{l}_2((j\psi)^*(P\nu),(j\psi)^*(Q\nu))=(j\psi)^*(\tilde{l}_2(P\nu,Q\nu))
\end{equation} for all $P, Q \in Loc_E$ if and only if $$\omega
^{ab}\E_b((Q\circ j\psi)\textup{det}\psi _M)\E_a((P\circ j\psi)
\textup{det}\psi _M) =[(\tilde{\omega}
^{ab}\tilde{\E}_b(Q)\tilde{\E}_a(P))\circ j\psi] (\textup{det}\psi _M),$$ for
all $P, Q \in Loc_E$, which by Lemma $\ref {Lagrangian}$ is equivalent to \\

$\displaystyle \omega ^{ab}(\textup{det}\psi _M)^2 \frac{\partial \psi^d}
{\partial u^b} \frac{\partial \psi^c}{\partial u^a}(\tilde{\E}_d(Q) \circ
j\psi)
(\tilde{\E}_c(P) \circ j\psi) = $
$$(\tilde{\omega} ^{ab}\circ j\psi)(\tilde{\E}_b(Q)\circ j\psi)
(\tilde{\E}_a(P)
\circ j\psi) (\textup{det}\psi _M).$$
The last equation is true for all $P, Q \in Loc_E$, so it is equivalent to
$$\displaystyle \tilde{\omega}^{ab}\circ j\psi = (\textup{det}\psi _M)
\omega ^{cd}\frac{\partial \psi_E^a}{\partial u^c} \frac{\partial
\psi_E^b}{\partial u^d}$$ which in turn is equivalent to the covariance of
$\omega$. \\ \\ \noindent
Now consider $l_3$ in degree 0. We have
\begin{eqnarray*}
l_3((j\psi)^* (P\nu),(j\psi)^* (Q\nu),(j\psi)^* (R\nu)) & = &
s[\tilde{l}_2(\tilde{l}_2 ((j\psi)^*(P\nu),(j\psi)^*(Q\nu)),(j\psi)^* \\ &
&
(R\nu))]\\
& = &
s[\tilde{l}_2((j\psi)^*(\tilde{l}_2(P\nu,Q\nu)),(j\psi)^*(R\nu))] \\ & = &
s[(j\psi)^*(\tilde{l}_2(\tilde{l}_2(P\nu,Q\nu),R\nu))] \\ & = &
s[-(j\psi)^*(l_1l_3(P\nu,Q\nu,R\nu))]\\
& = &
s[-l_1((j\psi)^*l_3(P\nu,Q\nu,R\nu))]
\end{eqnarray*}
since in this case $l_1=d_H$ so it
commutes with the pull-back (Proposition $\ref {commute}$). Proceeding
using the identity $$-s \circ l_1=1 + l_1 \circ s$$ the above becomes
$(1 + l_1 \circ s)[(j\psi)^*(l_3(P\nu,Q\nu,R\nu))] = $
$$(j\psi)^*(l_3(P\nu,Q\nu,R\nu)) +l_1 \circ
s[(j\psi)^*(l_3(P\nu,Q\nu,R\nu))].$$
So $l_3((j\psi)^*(P\nu),(j\psi)^*(Q\nu),(j\psi)^*(R\nu))=
(j\psi)^*(l_3(P\nu,Q\nu,R\nu))$ up to an exact form. We have shown:

\begin{theorem}
Let $\psi:E \rightarrow E$ be an automorphism of $E$ sending fibers to
fibers, and let $j\psi:J^\infty E \to J^\infty E$ be its
induced automorphism on $J^\infty E$. Then
$$\tilde{l}_2((j\psi)^* \alpha,(j\psi)^* \beta) =
(j\psi)^*(\tilde{l}_2(\alpha,\beta))$$
for all $\alpha,\beta \in \Omega^{n,0}(J^{\infty}E)$ iff $\omega$ is
covariant with respect to $\psi$. Moreover we then have
$$l_3((j\psi)^*\alpha,(j\psi)^*\beta,(j\psi)^*\gamma) =
(j\psi)^*l_3(\alpha,\beta,\gamma) + l_1(\delta),$$ for all
$\alpha,\beta,\gamma \in \Omega^{n,0}(J^{\infty}E)$, and for some $\delta
\in \Omega^{n-2,0}(J^{\infty}E)$.

\end{theorem}

\section{Reduction of the graded vector space in field theory} 

Let $M$ be a manifold, $E \rightarrow M$ a vector bundle, and $J^\infty E$ the
infinite jet bundle of $E$. Let $G$ be a Lie group acting on $E$ via
automorphisms (as in section $\ref{BAS}$) and hence inducing an action of $G$
on $J^\infty E$. We assume the induced action $\hat{\psi}_g$ on $\Gamma E$ is
canonical with respect to the Poisson bracket of local functionals for all
$g \in G$.
Notice that $G$ acts via canonical tranformations on the space of functionals
iff for every $j\psi _g$ $$\tilde{l}_2((j\psi _g)^* f_1,(j\psi _g)^*
f_2)=(j\psi _g)^* (\tilde{l}_2(f_1,f_2)),$$ where $\tilde{l}_2$ is defined on
the vector space $\Omega^{n,0}(J^\infty E)$ as in the previous section (in fact
$\displaystyle \tilde{l}_2(f_1,f_2) = \frac{1}{2} [\omega(\E(f_1),\E(f_2)) -
\omega(\E(f_2),\E(f_1))]$, see also equation $\ref {l2}$).

\begin{definition}
Given an automorphism $\psi$ of the bundle $E$, a differential form
$\alpha \in
\Omega ^{k,l}(J^\infty E)$ is {\em $\psi$-invariant} iff
$(j\psi)^*\alpha=\alpha$. If $G$ acts on $E$ via automorphisms $\psi _g:E
\rightarrow E, g \in G$, then $\alpha$ is {\em $G$-invariant} iff it is
$\psi
_g$-invariant for all $g \in G$.
\end{definition}
Let $\Omega ^{k,l}_\psi(J^\infty E)$ denote the space of all
$\psi$-invariant
forms on $J^\infty E$ which are in $\Omega ^{k,l}_c(J^\infty E)$, and let
$\Omega ^{k,l}_G(J^\infty E)$ denote the space of all $G$-invariant forms
in $\Omega ^{k,l}_c(J^\infty E)$. \\

\noindent One also needs the following:
\begin{definition}
Assume that $G$ acts on $E$ such that $J^\infty E/G$ has a manifold
structure
and the canonical projection map $\pi:J^\infty E \rightarrow J^\infty E/G$
is
smooth. Then
$\Omega ^{k,0}_c(J^\infty E/G)$ is the subspace of $k$-forms $\alpha \in
\Omega ^k_c(J^\infty E/G)$ such that $\pi ^* \alpha \in \Omega ^{k,0}_G
(J^\infty E)$, and
$\Omega ^{*,0}_c(J^\infty E/G)$ is the {\em reduced graded space} of the
graded space $\Omega ^{*,0}_c(J^\infty E)$ with respect to $G$.
\end{definition}

{\em We are interested in actions that send fibers to fibers, i.e. the
transformation of the independent variables does not depend on the
dependent
variables, so that acting on an element of $\Omega^{k,0}_c(J^\infty E)$
gives an
element of the same space and the reduction to $\Omega^{*,0}_c(J^\infty
E/G)$
makes sense.} \\

In fact we will also assume that the map $\psi_M$ representing the
transformation of the independent variables ($x^\nu = \psi ^\nu_M(x^\mu)$) is
the identity for all $g \in G$ (see the proposition that follows). This will
enable us to define a differential on the reduced graded space. It will also
insure that the space $\Omega^{n,0}_c(J^\infty E)$ does not collapse to zero
upon reduction (due to a reduction in the number of independent variables so
that any $n$-form in $\Omega ^{n,0}_c (J^\infty E /G)$ would be trivial) which
is desired so that the induced sh-Lie structure would not necessarily be
trivial.

\begin{proposition}
If $\psi _M$ is the identity map, then $\pi ^*:\Omega^{k,0}_c(J^\infty E/G)
\to
\Omega^{k,0}_G (J^\infty E)$ is onto.
\end{proposition}
\begin{proof}
Notice that if $\psi _M$ is the identity map one can choose coordinates
$\{x^i\}$ on $J^\infty E /G$ such that $\pi ^* x^i = x^i, i=1,2,\cdots,n$
(where by an abuse of notation we denote by $x^i$ both coordinates on $M$ and
$J^ \infty E /G$)
are
the coordinates on $M$. Now $\pi ^*(\alpha _I dx^I) = (\alpha _I \circ
\pi) dx^I$ where $|I|=k$, i.e. we are assuming
$\alpha _I dx^I \in \Omega^{k,0}_c(J^\infty E/G)$.
Since $\pi$ is a smooth canonical projection, it is clear that for any
smooth
$G$-invariant function $f$ on $J^\infty E$ there exists a smooth function
$\alpha _I$ on $J^\infty E /G$ such that $f=\alpha _I \circ \pi$. The
result
follows. \qed
\end{proof}
\begin{corollary} \label {isom}
If $\psi _M$ is the identity map, then we have an isomorphism $\pi ^*:
\Omega^{k,0}_c(J^\infty E/G) \longrightarrow \Omega^{k,0}_G (J^\infty E)$.
\end{corollary}

In this setting it can be shown that $\Omega^{*,0}_c(J^\infty E/G)$ is a
complex with a differential $\hat{d}_H:\Omega ^{m,0}_c(J^\infty E/G)
\longrightarrow \Omega ^{m+1,0}_c(J^\infty E/G)$ defined by $$\hat{d}_Hh
= (\pi ^*)^{-1}(d_H(\pi ^*h)).$$
This is well defined since $d_H(\pi ^*h)$ is invariant under the group
action
which follows from the fact that $\pi ^*h$ is invariant under the group
action
so that $(j\psi _g)^*(d_H(\pi ^*h)) = d_H((j\psi _g)^*(\pi ^*h)) = d_H(\pi
^*h)$. Also notice that $\hat{d}_H \circ \hat{d}_H =0$ easily follows from $d_H
\circ d_H=0$.
So $\hat{d}_H$ is a well-defined differential.
\\

\noindent {\bf Reduction hypothesis}: {\em Assume that every invariant
$d_H$-exact form is the horizontal differential of an invariant form. This 
hypothesis will guarantee that the reduced graded space with the differential
$\hat{d}_H$ is exact. Subsequently we will determine sufficient conditions
which will insure that this is true.} \\

This assumption will also yield the two conditions, $(i)$ and $(ii)$ below,
that are needed to obtain the sh-Lie structure on the reduced graded space.

\begin{lemma} \label {exact} Suppose that $\Omega ^{*,0}_c(J^\infty E)$ is
exact. If for every $d_H$-exact form $\alpha \in \Omega ^{k,0}_G(J^\infty E)$
there exists $\gamma \in \Omega ^{k-1,0}_G(J^\infty E)$ such that $\alpha = d_H
\gamma$, then the reduced graded space is exact. \end{lemma}
\begin{proof}
Suppose that $\hat{d}_H \beta = 0$, then $\pi ^*(\hat{d}_H \beta)=0$ and
by the
definition of $\hat{d}_H$ this implies that $d_H(\pi ^* \beta)=0$. Now
exactness
of $\Omega^{*,0}_c$ implies that $\pi ^* \beta = d_H \gamma$ for some
$\gamma$,
and $\gamma$ can be chosen so that it is invariant by assumption since
$d_H
\gamma$
is, so $\gamma = \pi ^* \tau$ for some $\tau$. By the definition
of $\hat{d}_H$ then $ d_H(\pi ^* \tau)=\pi ^*(\hat{d}_H\tau)$ so that $\pi
^*
\beta = \pi ^*(\hat{d}_H\tau)$ or $$\pi ^*(\beta - \hat{d}_H\tau)=0$$ from
which
$\beta - \hat{d}_H\tau=0$, and therefore $\beta = \hat{d}_H\tau$. ({\em
Observe that
$\pi ^* \delta = \delta \circ d\pi = 0$ implies that $\delta
= 0$ for $\delta \in \Omega ^{k,0}_c(J^\infty E/G)$ since $d\pi$ is onto}.)
\qed
\end{proof}

\begin{remark} We have used the simplified notation $\Omega^{*,0}_c$ for
$\Omega^{*,0}_c (J^\infty E)$. \end{remark}

\begin{corollary} \label {exactc}
Under the same hypothesis as in the preceeding lemma, the subcomplex of
$G$-invariant forms,
$\Omega ^{*,0}_G(J^\infty E)$, is exact. \end{corollary}

Now we proceed to finding a map on the reduced space $\Omega^{n,0}_c(J^\infty
E/G)$ analogous to and induced by
$\tilde{l}_2$ on the space $\Omega^{n,0}_c(J^\infty E)$. Define
$\hat{l}_2$ by $$(\hat{l}_2(f_1,f_2)) =
(\pi ^*)^{-1}\tilde{l}_2(\pi ^*f_1,\pi ^*f_2)$$ where $f_1,f_2
\in \Omega^{n,0}_c(J^\infty E/G)$. Notice
that this is well-defined since $\tilde{l}_2(\pi ^*f,\pi ^*h)$ is
invariant
under the group action by the following
calculation $$(j\psi _g)^*\tilde{l}_2(\pi ^*f,\pi ^*h) =
\tilde{l}_2((j\psi _g)^*(\pi ^*f),(j\psi _g)^*(\pi ^*h)) =
\tilde{l}_2(\pi ^*f,\pi ^*h),$$ and the map $(\pi ^*)^{-1}$ exists by Corollary
$\ref {isom}$.

Skew-symmetry and linearity of
$\hat{l}_2$ follow from the skew-symmetry and linearity of
$\tilde{l}_2$.
Furthermore $\hat{l}_2$ satisfies \\

$(i) \quad \hat{l}_2(\hat{d}_H k_1,h) = \hat{d}_H k_2,$ \\ \indent
$\displaystyle (ii) \quad \sum _{\sigma \in unsh(2,1)} (-1)^\sigma
\hat{l}_2(\hat{l}_2(f_{\sigma(1)},f_{\sigma(2)}),f_{\sigma(3)}) =\hat{d}_H
k_3,$
\\

\noindent
where $k_1 \in \Omega ^{n-1,0}_c(J^ \infty E/G)$, while $h,f_1,f_2,f_3 \in
\Omega ^{n,0}_c(J^ \infty E/G)$, and for some $k_2,
k_3 \in \Omega ^{n-1,0}_c(J^ \infty E/G)$. Recall that we may suppress
some of the notation and assume
the summands are over the appropriate shuffles with their corresponding
signs.
\\ \\ \indent
To verify $(i)$ notice that
\begin{eqnarray*}
\pi ^*(\hat{l}_2(\hat{d}_Hk_1,h))
& = &
\tilde{l}_2(\pi ^*(\hat{d}_Hk_1),\pi ^*h) \\ & = &
\tilde{l}_2(d_H(\pi ^*k_1),\pi ^*h) \\
& = &
d_H K_2,
\end{eqnarray*}
for some $K_2 \in \Omega^{n-1,0}_c(J^ \infty E)$. But by our assumption
$K_2$
can be
chosen to be invariant under the group action since $d_H K_2$ is, i.e.,
$K_2=\pi
^*k_2$ for some $k_2 \in \Omega ^{n-1,0}_c(J^\infty E/G)$, and then $d_H K_2
= d_H(\pi ^*k_2) = \pi ^*(\hat{d}_H k_2)$ by the definition of
$\hat{d}_H$.
This implies that $\hat{l}_2(\hat{d}_Hk_1,h) = \hat{d}_Hk_2$. ({\em Recall
that $\pi ^* \alpha = \alpha \circ d\pi = 0$ implies that $\alpha
= 0$ for $\alpha \in \Omega ^{n,0}_c(J^\infty E/G)$ since $d\pi$ is onto}.) \\

While to verify $(ii)$, notice that 

\begin{eqnarray*}
\displaystyle \pi ^*(
\hat{l}_2(\hat{l}_2(f_1,f_2),f_3))
& = &
\tilde{l}_2(\tilde{l}_2(\pi
^*f_1,\pi ^*f_2),\pi ^*f_3) \\
& = &
d_H K_3,
\end{eqnarray*}
where the sum is over the unshuffles (2,1), and for some $K_3 \in
\Omega^{n-1,0}_c(J^\infty E)$ and all $f_1,f_2,f_3 \in \Omega^{n,0}_c(J^\infty
E /G)$. Again $K_3$ can be chosen
to be invariant under the group action since $d_H K_3$ is, i.e., $K_3=\pi
^*k_3$ for some $k_3 \in \Omega ^{n-1,0}_c(J^ \infty E/G)$, and then $d_H
K_3
= d_H(\pi ^*k_3) = \pi ^*(\hat{d}_H k_3)$ by the definition of
$\hat{d}_H$.
This implies that
$\displaystyle \hat{l}_2(\hat{l}_2 (f_1,
f_2),f_3) = \hat{d}_Hk_3$. \\ \\
\noindent
We have shown (see Lemmas 1 and 2 in \cite{BFLS98} and Theorem $\ref{shLie}$):

\begin{theorem}
There exists a skew-symmetric bilinear bracket on $H_0(\hat{d}_H) \times
H_0(\hat{d}_H)$ that satisfies the Jacobi identity, where we are using
$H_0(\hat{d}_H)$ for $H^n(\Omega ^{*,0}_c(J^\infty E/G),\hat{d}_H)$.
This bracket is induced by the map $\hat{l}_2$.
\end{theorem}

\begin{theorem}
The skew-symmetric linear map $\hat{l}_2$ as defined above on the space
$\Omega ^{n,0}_c(J^\infty E/G)$ extends to an sh Lie structure on the graded
space
\\ $\Omega ^{*,0}_c(J^\infty E/G)$.
\end{theorem}

\subsection{Exactness of the reduced graded space} In this section we find
sufficient conditions under which our reduction hypothesis in the last
section holds. Thus we consider the question: If $\alpha$
is in the reduced space $\Omega ^{k-1,0}_c(J^ \infty E/G)$ and $\hat{d}_H
\alpha =
0$, then is $\alpha = \hat{d}_H \beta$ for some $\beta$? Suppose
$\hat{d}_H
\alpha = 0$ for $\alpha \in \Omega ^{k,0}_c(J^ \infty E/G)$,
then $d_H(\pi ^* \alpha) = 0$ so that $\pi ^*\alpha = d_H \gamma$ for some
$\gamma \in \Omega^{k-1,0}_c(J^ \infty E)$ since $\Omega^{*,0}_c(J^ \infty
E)$
is exact. Notice that $d_H \gamma$ is invariant under the group action
(since
$d_H \gamma = \pi ^*\alpha$) so $d_H \gamma = j\psi _g^*(d_H \gamma)$ for
all $g \in G$, or
since $d_H$ commutes with $j\psi _g^*$ by proposition $\ref {commute}$,
$d_H
\gamma = d_H(j\psi _g^* \gamma)$ for all $g \in G$. So $$\gamma = j\psi
_g^* \gamma + d_H \tau _g,$$ where $\tau _g \in \Omega ^{k-2,0}_c(J^
\infty
E)$ depends on $g$. Consider $\gamma ' = \gamma + d_H \Delta$ for some
fixed $\Delta \in \Omega
^{k-2,0}_c(J^\infty E)$ and notice that $d_H \gamma ' = d_H \gamma = \pi
^*
\alpha$.
Now $\gamma ' = \gamma + d_H \Delta = j\psi _g^* \gamma + d_H \tau _g +
d_H
\Delta$ so
that $j\psi _g^* \gamma = \gamma ' - d_H \tau _g - d_H \Delta$, and hence
$j\psi _g^* \gamma ' = j\psi _g^* \gamma + j\psi _g^* (d_H \Delta) =
\gamma
'
- d_H \tau _g - d_H \Delta + j\psi _g^* (d_H \Delta)$. But if $\gamma '$
is
invariant under the group action then $- d_H \Delta - d_H
\tau _g + j\psi _g^* (d_H \Delta) = 0$ or $$d_H(j\psi _g^* \Delta - \Delta
-
\tau _g) = 0,$$
(recall that $d_H$ commutes with $j\psi _g^*$ by proposition
$\ref{commute}$) so
that $(j\psi _g^* \Delta - \Delta - \tau _g)$ is exact. Notice that this is a
necessary and sufficient condition for the exactness of the reduced space.
In this case let $\beta=
\pi _*\gamma '$ and notice that $\pi ^* (\hat{d}_H \beta)= d_H \gamma '=
\pi ^* \alpha$ so that $\hat{d}_H \beta = \alpha$. \\ \\ \noindent
Observe
that $\tau _g$ depends on $g$ and on $\gamma$ whereas $\Delta$
depends on $\gamma$. \\

We find the above criterion too general and rather complicated, and find it
useful to consider a special case. Suppose that $G$ is {\em compact} and let
$\alpha \in \Omega ^{k,0}_c(J^ \infty E)$ be a closed form that is invariant
under the group action. By exactness of $\Omega ^{*,0}$ there exists a $\beta$
such that $d_H \beta = \alpha$. Observe that $d_H (j\psi _g^* \beta) =j\psi
_g^* (d_H \beta) = j\psi _g^* \alpha = \alpha$ for all $g \in G$. So \[ \int_G
d_H (j\psi _g^* \beta) dg = \int_G \alpha dg = \alpha \int_G dg = \alpha \cdot
vol(G) = \alpha \] assuming that $vol(G) =1$. Now let \[ \hat{\beta} = \int_G
(j\psi _g^* \beta) dg \] and notice that $\displaystyle d_H \hat{\beta} =
\int_G d_H (j\psi _g^* \beta) dg = \alpha$, and $\displaystyle j\psi _h^*
(\hat{\beta}) = \int_G j\psi _h^*(j\psi _g^* \beta) dg = \int_G (j\psi _{gh}^*
\beta) d(gh) = \hat{\beta}$. So we have:

\begin {proposition} If the group $G$ acting (canonically) on $E$ is compact,
then every $d_H$-closed form that is $G$-invariant is the horizontal 
differential of a $G$-invariant form. Consequently $\Omega ^{*,0}_c(J^\infty
E/G)$ is exact and admits an (induced) sh-Lie structure. \end {proposition}

\subsection{ The existence of an sh-Lie structure on the subcomplex of
$G$-invariant forms} In this subsection we consider the subcomplex of
$G$-invariant forms $$\cdots \rightarrow \Omega ^{n-1,0}_G(J^\infty E)
\xrightarrow{d_H} \Omega ^{n,0}_G(J^\infty E).$$ Working with the subcomplex of
$G$-invariant forms is rather interesting. We shall maintain the same
assumptions made earlier in this section, in particular the hypotheses of Lemma
$\ref {exact}$. Recall that throughout this section we require the mapping
$\psi _M$ representing the tranformation of the independent variables to be the
identity.

In fact, if the base manifold $M$ is one-dimensional these assumptions are not
needed for this subcomplex to be exact. However their absence does not
guarantee the existence of an sh-Lie structure (that's obtained >from the
original one). \\ \indent Recall that by Corollary $\ref {exactc}$ the
subcomplex of $G$-invariant forms is exact and observe that $\tilde{l}_2 (
\alpha,\beta)= \tilde{l}_2((j\psi)^*\alpha, (j\psi)^*\beta) =  (j\psi)^*
\tilde{l}_2(\alpha,\beta)$ for $\alpha , \beta \in \Omega ^{n,0}_G (J^\infty
E)$. So $\tilde{l}_2$ can be restricted to the subspace $\Omega
^{n,0}_G(J^\infty E)$, and if we combine this with exactness, we notice that
conditions $(i)$ and $(ii)$ that guarantee the existence of the sh-Lie
structure, as stated earlier in this section, are readily established (this
sh-Lie structure is just the restriction of the original one to the subcomplex
of $G$-invariant forms). So we have:

\begin{theorem}
Under the same hypotheses as in Lemma $\ref {exact}$, there exists an sh-Lie
structure on the subcomplex of $G$-invariant forms $\Omega ^{*,0}_G(J^\infty
E)$. \end{theorem}

\begin {example} Consider $M={\mathbf R}, E={\mathbf
R} \times {\mathbf R}^2$, and let \[ \omega = \left( \begin{array}{cc} 0 & 1 \\
-1 & 0 \end{array} \right). \] Consider the action of $G= SO(2)$ on $E$ defined
by $\psi _g(x,u^1,u^2) = (x, g\cdot (u^1,u^2)), g \in SO(2)$, and observe that
$SO(2)$ is compact. As we noticed in the example in section 1, $\omega$ is
covariant with respect to $\psi_g$ and hence the induced action is canonical
for all $g \in G$. Consider a subset of $J^\infty E$ defined by $(J^\infty E)'
= J^\infty E - \{u \in J^\infty E|u=(u^1 _I,u^2 _I)=(0,0), I=0,1,2,3, \cdots
\}$, where $u^1_0 = u^1, u^1_1 = u^1_x, u^1_2 = u^1_{xx}, \cdots$ etc.

Notice that at a point in $(J^\infty E)'$ $u^1$ and $u^2$ can't be zero at the
same time, $u^1_x$ and $u^2_x$ can't be zero at the same time, and so on.
We note that $(J^\infty E)' /G = \mathbf{R} \times (\mathbf{R}^+)
\times (\mathbf{R}^+) \times S^1 \times (\mathbf{R}^+) \times S^1 \times
\cdots$, where $\mathbf{R}^+$ is the set of all positive real numbers (without
the 0). As an illustration let $\alpha = u^1 dx$ and notice that $(j\psi _g)^*
\alpha = (\cos \theta) u^1 + (\sin \theta) u^2$, whereas $\beta = (1/2)[(u^1)^2
+ (u^2)^2] dx$ is invariant under the induced action of $G$ and so is $\gamma =
(1/2)[(u^1_x)^2 + (u^2_x)^2] dx$. As for the sh-Lie structure, first notice
that the resolution space is rather simple: $$0 \to Loc_E^0 \to \Omega
^{1,0}_{c}(J^\infty E).$$ One finds that $$\tilde{l}_2(Pdx, Qdx) =
(\E_1(P)\E_2(Q) - \E_2(P)\E_1(Q)) dx,$$ while $l_2(Pdx, f)=0$ for $f \in
Loc_E^0$ since $l_2l_1(Pdx, f)=l_2(l_1(Pdx),f) + l_2(Pdx,l_1 f) = 0+l_2(Pdx,
d_H f)=0$ which follows from $\E_i(d_H f)=0, i=1,2$. Further $l_2 = 0$ in higher
degrees. We note that $l_3$ is non-zero on degree 0, but is zero on higher
degrees. Let for example $P_1=u^1 u^2_x, P_2=u^1u^2$, and $P_3=(u^1)^2$, then
$\tilde{l}_2\tilde{l}_2(P_1dx, P_2dx, P_3dx) = 4u^1u^1_x dx = d_H(2(u^1)^2),$
so that we can choose $l_3(P_1dx, P_2dx, P_3dx) = -2(u^1)^2$.

The subcomplex of $G$-invariant forms is exact and so the sh-Lie structure can
be restricted to it. Observe that, for example, $\tilde{l}_2(\beta, \gamma) =
\tilde{l}_2((1/2)[(u^1)^2 + (u^2)^2] dx, (1/2)[(u^1_x)^2 + (u^2_x)^2] dx) =
(-u^1u^2_{xx} + u^2u^1_{xx}) dx$ is invariant under the induced group action of
$G=SO(2)$. \qed \end {example}

\begin{remark} In \cite{KO03} Kogan and Olver provide a definitive account of
invariant Euler-Lagrange equations using moving frames and other tools from
differential geometry. We note that their invariantization map $\imath$ in our
case will just map horizontal differentials to themselves (i.e. in local
coordinates $\imath (dx^i) = dx^i$). Consequently the invariant derivatives are
the same as the ordinary derivatives $\mathcal D _i = D_i$ (here we are
borrowing some of the notation from \cite{KO03}) whereas the twisted  invariant
adjoints of $\mathcal D _i$ turn out to be $\mathcal D^\dagger _i = -\mathcal
D_i$. Now suppose that $I^1, \cdots, I^m$ is a fundamental set of differential
invariants (on $J^\infty E$). Let $\hat{I}^1, \cdots, \hat{I}^m$ be coordinates
on $J^\infty E /G$ such that $\hat{I}^k\circ \pi = I^k, k=1, \cdots,m$ where
$\pi$ is the canonical projection map $\pi:J^\infty E \to J^\infty E /G$, and
let $\hat{L}$ be the corresponding Lagrangian defined on $J^\infty E /G$ for
some (invariant) Lagrangian $L = \hat{L}\circ \pi$ defined on $J^\infty E$. The
{\em invariant Eulerian} (in \cite{KO03}) is defined by $$\mathcal E_\alpha
(\tilde{L})= \sum _K \mathcal D_K^{\dagger} \frac{\partial \tilde{L}}{\partial
I^{\alpha}_{,K}}$$ where $\tilde{L}$ indicates the Lagrangian is written in
terms of the $I^{\alpha}_{,K}$'s. But this $\mathcal E_\alpha$, in our case,
reduces to $$\mathcal E_\alpha (\tilde{L})= \sum _K (-D)_K \frac{\partial
\tilde{L}} {\partial I^{\alpha}_{,K}}.$$ A {\em corresponding invariant
Eulerian} can be defined on $J^\infty E /G$. To accomplish this, first define
``total derivative" $\hat{D}_k$ on $J^\infty E/G$ by $(\hat{D}_k \hat{P})\circ
\pi = D_k(\hat{P} \circ \pi)$ where $\hat{P}$ is a smooth function on $J^\infty
E/G$. Now let $\hat{I}^{\alpha}_{,K} = \hat{D}_K \hat{I}^{\alpha}$ and notice
that $\hat{I}^{\alpha}_{,K} \circ \pi= I^{\alpha}_{,K} = D_K I^{\alpha}$.
Finally, define $$\hat{\mathcal E}_\alpha (\hat{L})= \sum _K \mathcal
(-\hat{D})_K \frac{\partial \hat{L}}{\partial \hat{I}^{\alpha}_{,K}}$$ and
observe that $\hat{\mathcal E}_\alpha (\hat{L}) \circ \pi = \mathcal E_\alpha
(\tilde{L})$. The reader should consult \cite{KO03} for more details.
\end{remark}

\section{Functional invariance}
In this section we consider the implications of invariance on the space of
functionals. Assume that $\psi$ is canonical, i.e. $$\{\mathcal P\circ
\hat{\psi},\mathcal Q \circ \hat{\psi} \}= \{\mathcal P,\mathcal Q\} \circ
\hat{\psi}$$ for all functionals $\mathcal P,\mathcal Q$.
\begin{definition} We say that ${\mathcal P}$ is invariant under $\psi$ iff
${\mathcal P}\circ \hat{\psi} = {\mathcal P}$. \end{definition}
Notice that this holds if, and only if $$\int_M(P\circ j\psi\circ
j\phi)\textup{det}\psi _M \nu = \int_M(P\circ j\phi) \nu$$ for all $\phi \in
\Gamma E$, which in turn holds if $(P\circ j\psi)\textup{det} \psi _M =P$.
Similarly one says that $P \in Loc_E^0$ is invariant under $\psi$ iff $(P\circ
j\psi) \textup{det} \psi _M = P$. Let ${\mathcal F}_\psi$ denote the set of all
functionals ${\mathcal P}$ such that ${\mathcal P}\circ \hat{\psi} = {\mathcal
P}$. Observe that $${\mathcal P,Q} \in {\mathcal F}_\psi \Rightarrow
\{{\mathcal P,Q}\} \in {\mathcal F}_\psi$$ so ${\mathcal F}_\psi$ is a Lie
subalgebra of ${\mathcal F}$. Let $Loc_E^0(\psi)$ denote the subset of
$Loc_E^0$ consisting of $P \in Loc_E^0$ such that $$P = (P\circ
j\psi)\textup{det}\psi _M.$$
We note that $Loc_E^0(\psi)$ is a subspace of $Loc_E^0$, while for
automorphisms $\psi$ such that $\textup{det}\psi _M=1, Loc_E^0(\psi)$ is a
subalgebra of $Loc_E^0$.

\begin{proposition} \label{EI}
If $P\nu \in \Omega ^{n,0}(J^\infty E)$ is $\psi$-invariant for an
automorphism
$\psi$, then so is $\E(P\nu)$.
\end{proposition}
\begin{proof}
In local coordinates $\E(P\nu)=\E_a(P)(\theta ^a \wedge \nu)$. So
\begin{eqnarray*}
(j\psi)^*(\E(P\nu))
& = &
(j\psi)^*(\E_a(P))(j\psi)^*(\theta ^a \wedge \nu) \\ & = &
(\E_a(P)\circ j\psi)(\frac{\partial \psi ^a_E}{\partial u^b} \theta ^b
\wedge
(\textup{det}\psi _M) \nu) \\
& = &
(\E_a(P)\circ j\psi)\frac{\partial \psi ^a_E}{\partial u^b}
(\textup{det}\psi _M)
(\theta ^b \wedge \nu).
\end{eqnarray*}
Now, since $P\nu$ is $\psi$-invariant we have
\begin{eqnarray*}
P\nu
& = &
(j\psi)^*(P\nu) \\
& = &
(j\psi)^*(P)(j\psi)^*\nu \\
& = &
(P\circ j\psi)(\textup{det}\psi _M) \nu,
\end{eqnarray*}
and therefore $(P\circ j\psi)\textup{det}\psi _M = P$. Finally,
\begin{eqnarray*}
\E(P\nu)
& = &
\E_a(P)(\theta ^a \wedge \nu) \\
& = &
\E_a((P\circ j\psi)\textup{det}\psi _M)(\theta ^a \wedge \nu) \\ & = &
(\textup{det}\psi _M)\frac{\partial \psi ^b_E}{\partial u^a} (\E_b(P)\circ
j\psi)
(\theta ^a \wedge \nu) \\
& = &
(j\psi)^*(\E(P\nu)),
\end{eqnarray*}
where we have used Lemma $\ref {Lagrangian}$ in the last calculation. \qed
\end{proof}

If $G$ is a Lie group which acts on $E$ via canonical automorphisms
$\psi_g$, for all $g\in G,$  then we write
$$Loc_E^0(G) =\bigcap_{g\in G}Loc_E^0(\psi_g) \quad ,\quad
{\cal F}_G=\bigcap_{g\in G}{\cal F}_{\psi_g}.$$
Clearly ${\cal F}_G$ is a Lie sub-algebra of ${\cal F}$, and if
$P\in Loc_E^0(G)$, then $\E(P\nu)$ is $G$-invariant.

Notice that if $\phi$
is a section of the bundle $E\rightarrow M$ then $j^{\infty}\phi$ is a
section of
$\pi^{\infty}:J^{\infty}E\rightarrow M.$ Sections of this type are said to
be {\bf
holonomic} as they are induced by a section of $E\rightarrow M.$ It is
easily shown that not
all sections of $\pi^{\infty}$ are holonomic. Observe that $\pi\circ
j^{\infty}\phi$ is a
section of the bundle $\tau:J^{\infty}E/G\rightarrow M$ since
$\pi^{\infty}=\tau \circ \pi.$
Similarly we say that a section $\eta$ of $\tau$ is {\bf holonomic} if it
has the form
$\eta=\pi\circ j^{\infty}\phi$ for some section $j^{\infty}\phi$ of
$\pi^{\infty}.$
Let $\Gamma$
denote the set of all holonomic sections of $\tau.$ Note that $\Gamma$ is
not a linear
space over ${\bf R}$ since $\pi$ is not linear. Indeed $J^{\infty}E/G$ is
generally not
a vector bundle.

\begin{definition} We say that $\tilde {\cal P}$ is a
{\it reduced local functional} if it is a mapping from the set $\Gamma$ of
holonomic sections of the bundle $\tau:J^{\infty}E/G \rightarrow M$
into ${\bf R}$ such that
$$\tilde {\cal P}(\eta)=\int_M \eta^*(\tilde P) \nu$$ for some smooth
mapping $\tilde P:J^{\infty}E/G \rightarrow {\bf R}$ and for every $\eta\in
\Gamma.$ We
denote the set of all reduced local functionals by $\tilde {\cal F}.$
\end{definition}

In this definition, when we say that $\tilde P:J^{\infty}E/G \rightarrow {\bf
R}$ is smooth we mean that $\tilde P\circ \pi$ is in $Loc_E^0.$

\begin{proposition} \label{FI} There is a bijection $\Xi$
from $\tilde {\cal F}$ onto ${\cal F}_G.$ The mapping $\Xi$ is defined
as follows: if $\tilde {\cal P}$ is defined by
$$\tilde {\cal P}(\eta)=\int_M \eta^*(\tilde P) \nu$$ for some smooth
mapping $\tilde P,$
then $\Xi(\tilde {\cal P})={\cal P}$ is defined by
$${\cal P}(\phi)=\int_M (j^{\infty}\phi)^*( P) \nu$$ where $P=\tilde P\circ
\pi.$
\end{proposition}

The proof of the proposition is straightforward and is omitted.

\begin{remark} It follows from the proposition that the set $\tilde {\cal F}$
of reduced local functionals inherits a Lie-structure from that on ${\cal
F}_G.$ In the sequel it is identified with ${\cal F}_G$. \end{remark}

Notice that by proposition $\ref {EI}$ the complex $$\Omega ^{0,0}_c
\xrightarrow{d_H} \Omega ^{1,0}_c\xrightarrow{d_H} \cdots
\xrightarrow {d_H} \Omega
^{n-1,0}_c\xrightarrow {d_H} \Omega ^{n,0}_{c}\xrightarrow {\E} \Omega
^{n,1}_c
\rightarrow \cdots$$
induces a subcomplex
$$\Omega ^{0,0}_\psi \xrightarrow{d_H}\Omega ^{1,0}_\psi\xrightarrow{d_H}
\cdots
\xrightarrow{d_H}\Omega ^{n-1,0}_\psi\xrightarrow{d_H} \Omega ^{n,0}_\psi
\xrightarrow{\E} \Omega ^{n,1}_\psi\rightarrow \cdots$$ This
subcomplex is exact up to the term $\Omega ^{n-1,0}_\psi$ (and including it,
i.e. $H^{n-1}_\psi = 0)$, if we assume that $\Omega _c^{*,0}$ is itself exact
and that every exact $\psi$-invariant form is the horizontal differential of
some $\psi$-invariant form. Similarly, the subcomplex $$\Omega ^{0,0}_G
\xrightarrow{d_H} \Omega ^{1,0}_G\xrightarrow{d_H} \cdots
\xrightarrow{d_H} \Omega ^{n-1,0}_G\xrightarrow{d_H} \Omega ^{n,0}_G
\xrightarrow{\E}
\Omega ^{n,1}_G\rightarrow \cdots$$ is exact up to the term $\Omega
^{n-1,0}_G$
(under the same assumptions).

\section{An example: A Poisson Sigma model} A number of authors
\cite{CF99,I94,SS94} have investigated a class of physical
theories called Poisson sigma models. These models focus on fields which are
defined on a 2-dimensional manifold $\Sigma$ with range in a Poisson manifold
$M.$ These models seem to have first arisen in various theories of
2-dimensional gravity but have been applied to areas such as topological field
theory and in the reformulation of Kontsevich's work on deformation
quantization \cite{CF99}. We consider an application of our results to the
version of the Poisson sigma model presented in the work of Ikeda \cite{I94}
but we utilize the notation of \cite{BFLS98}.

Assume that $V$ is a finite-dimensional vector space, say of dimension $N$ and
with basis $\{T_A\}$,
and let $\{T^A\}$ denote the basis of the space $V^*$ dual to $V$. We
assume the existence of a Poisson tensor $W$ on $V^*.$ Thus $W$
is a bivector field
$$W=
W_{AB}(\frac {\partial}{\partial T_A}\wedge \frac {\partial}{\partial T_B})$$
where, for each $A,$ $T_A$ is identified as a coordinate mapping
$T_A:V^* \rightarrow {\mathbf R}$ and $V$ is identified with $V^{**}.$ The
fact that $W$ is Poisson means that it is a tensor and the
$\{W_{AB}\}$ are functions on $V^*$
(assumed  to be polynomials in the coordinates $\{T_A\}$ in the present model)
such that
$$W_{AD}\frac {\partial W_{BC}}{\partial T_D}+
W_{BD}\frac {\partial W_{CA}}{\partial T_D}+
W_{CD}\frac {\partial W_{AB}}{\partial T_D}=0$$
and $W_{AB}=-W_{BA}.$
Now $V^*$ is a Poisson manifold with
$$\{f,g\}=
W_{AB}\frac {\partial f}{\partial T_A}
\frac {\partial g}{\partial T_B}$$
for smooth functions $f,g$ defined on $V^*.$

Observe that the Poisson field $W$ is not dependent on the basis
used to represent it. If the components $W_{AB}$ of $W$  relative
to a basis $\{T_A\}$ of $V$ satisfy the Poisson conditions given above then
the components $\overline W_{AB}$ of $W$ relative to any other basis
$\{\overline T_B\}$ of $V$ will also satisfy these same conditions.
Notice that $W_{AB}=\{T_A,T_B\}$ and that if $\{\overline T_A\}$ is
a different basis and $\{\overline W_{AB}\}$ are the components of
$W$ relative to it then $\overline W_{AB}=
\{\overline T_A,\overline T_B\}$ as well. \\

The fields of Ikeda's model are ordered pairs $(\psi,h)$ where $\psi$ is a
mapping from the 2-dimensional manifold $\Sigma$ into $V^*$ and $h$ is
a mapping from $\Sigma$ into $T^*\Sigma\otimes V.$ In components
$$\psi(x)=\psi_A(x)T^A \quad ,\quad h(x)=h^A_{\mu}(x)(dx^{\mu}\otimes T_A)$$
where $\{x^{\mu}\}$ are coordinates on $\Sigma.$
One form of Ikeda's Lagrangian for 2-dimensional gravity is
$$L= \epsilon^{\mu\nu}[h^A_{\mu}D_{\nu}\psi_A-
\frac{1}{2}W_{AB}h^A_{\mu}h^B_{\nu}],$$
where $\epsilon$ is the skew-symmetric Levi-Civita tensor on $\Sigma$
such that $\epsilon^{01}=1$  and
$$D_{\nu}\psi_A=\partial_{\nu}\psi_A+W_{AB}h^B_{\nu}.$$

It is our intent to show how some of our work relates to Ikeda's model.
To cast this model in our formalism let $E$ denote the vector
bundle over $\Sigma$ with total space $E=V^*\oplus [T^*\Sigma \otimes V]$
and with the obvious projection of $E$ onto $\Sigma.$ The fields $(\psi,h)$
are sections of this bundle. \\

First we show how to define a Poisson bracket
on the relevant space of local functionals. To accomplish this,
we want to construct a mapping $\omega$ as in section 2.1
in such a manner that the Jacobi condition is satisfied.

For this purpose we find it convenient to introduce a positive definite
metric $\mu$ on $V$ with its induced metric $\mu^*$ on $V^*.$ Moreover
we choose $\{T_A\}$ to be an orthonormal basis relative to $\mu$ and
we define $\{T^B\}$ by $T^B(v)=\mu(v,T_B)$ for each $B$ and for all
$v\in V.$ It follows that $\{T^A\}$ is a $\mu^*$-orthonormal basis of
$V^*$ and that the basis $\{T^A\}$ is dual to $\{T_B\}.$ Define a tensor
$\tilde W$ on $V$ by
$$\tilde W=
W^{AB}(\frac {\partial}{\partial T^A}\wedge \frac {\partial}{\partial T^B})$$
where $W^{AB}=\mu^{AC}\mu^{BD}W_{CD}$, and $\mu^{PQ}=\mu^*(T^P,T^Q)$ for $1 \leq
P,Q \leq N$.
Thus $\tilde {W}$ is the tensor
on $V$ induced by $W$ and the metric $\mu.$ \\

We reformulate this data in terms of the jet bundle of
$E$. In particular the tensors $w,\tilde w$ induce a bilinear mapping $\omega$
on the jet bundle which is used to define a Lie structure on the space of
functionals.
Local coordinates on $E$ may be denoted $(x^{\mu},u_A,w^B_{\mu})$
and those on the jet bundle $J^{\infty}E$ by $(x^{\mu},u_{A_I},w^B_{\mu,J})$.
Thus if $(\psi,h)$ is a section of $E$ we
have $$x^{\mu}((\psi,h)(p))=x^{\mu}(p) \quad ,\quad u_A((\psi,h)(p))=\psi_A(p)$$
and
$$w^B_{\mu}((\psi,h)(p))=h^B_{\mu}.$$
Clearly, there is a corresponding
splitting of the jet coordinates. It follows that in local coordinates
each local function $P$ on $J^{\infty}E$ is a function of
$(x^{\mu},u_{A,I},w_{\nu,J}^B).$ Now the $\{W_{AB}\}$ are functions of
the coordinates $\{T_A\}$ on $V$ and these coordinates are denoted
$\{u_A\}$ on the bundle $E.$ Consequently we can regard the
$\{W_{AB}\}$ as being functions on the jet bundle $J^{\infty}E$
which depend polynomially on the coordinates $\{u_A\}$ and are in fact
independent of the coordinates $x^{\mu},u_{A,I},w^B_{\mu,J}$ for $|I| \geq 1.$
The function $\omega$ is required to be a mapping from
$\Omega ^{n,1}_0 \times \Omega ^{n,1}_0$ into $Loc_E.$ Observe that
there are two types of contact forms $\theta_A, \theta^B_{\mu}$
on $J^{\infty}E$, those which arise
from the coordinates $\{u_A\}$ and those which arise from $\{w^B_{\mu}\}.$
Since each fiber of $E$ is a direct sum of two vector spaces the matrix of
components of $\omega$ is a block diagonal matrix with two blocks
defined by
$$\omega_{AB}=W_{AB} \quad \textup{and} \quad
\omega^{A,B}_{\mu,\nu}=\delta_{\mu\nu}W^{AB}.$$ Here $\delta_{\mu\nu}$
is the usual Kronecker delta symbol. In the first block
we have written the indices
of $\omega^{ab}$ as lower indices as they represent components
relative to a basis of the dual of $V.$  In the second block
it is appropriate to
write the components $\omega^{(\mu,A),(\nu,B)}$ as
$\omega_{\mu,\nu}^{A,B}$
for similar reasons. Notice that the matrix
of $\omega$ is skew-symmetric and
$$\omega_{AD}\frac {\partial \omega_{BC}}{\partial u_D}+
\omega_{BD}\frac {\partial \omega_{CA}}{\partial u_D}+
\omega_{CD}\frac {\partial \omega_{AB}}{\partial u_D}=0.$$
The other components of $\omega$ satisfy a similar conditon
as they too are determined by the $\{W_{AB}\}.$
It follows from this fact and equation (7.11) of \cite{O86} that the bracket
of local functionals defined on sections of $E$ by

$$\{{\cal P},{\cal Q}\}(\phi)=
\int_{\Sigma}[\omega^{ab}{\bf E}_a(P){\bf E}_b(Q)]\circ(j^\infty \phi) \nu$$
satifies the Jacobi identity (see the discussion in section 2.1). \\

Ikeda shows that the Euler operators for the Lagrangian $L$ in this model
are given by

$${\bf E}^A(L)=\epsilon^{\mu\nu}R^A_{\mu\nu}\quad ,\quad
 {\bf E}^{\mu}_A( L)=\epsilon^{\mu\nu}D_{\nu}\psi_A$$
where
$$R^A_{\mu\nu}=\partial_{\mu}h^A_{\nu}-\partial_{\nu}h^A_{\mu}+
\frac{\partial W_{BC}}{\partial T_A}h^B_{\mu}h^C_{\nu}.$$

\noindent
This suggests that for every local function $P$ we should define 
$${\bf E}^A(P)=(-D)_I(\frac {\partial P}{\partial u_{A,I}})
\quad ,\quad {\bf E}^{\mu}_B(P)=
(-D)_J(\frac {\partial P}{\partial w^B_{\mu,J}}).$$
Consequently, the Poisson bracket assumes the following
form: \\
\noindent
$\{{\cal P},{\cal Q}\}(\psi,h)=$ $$
\int_{\Sigma}[\omega_{AB}{\bf E}^A(P){\bf E}^B(Q)]
\circ j^{\infty}(\psi,h) \nu+
\int_{\Sigma}[\omega_{\mu,\nu}^{A,B}
{\bf E}^{\mu}_A(P)\E^{\nu}_B(Q)]\circ j^{\infty}(\psi,h) \nu.$$

Now we characterize the automorphisms of $E$ that induce canonical
transformations, relative to the Poisson bracket, on the space of functionals
${\mathcal F}$. Recall that in section 2.2 we have referred to such
automorphisms as canonical automorphisms.
Suppose that $\Psi$ is a linear automorphism of $E,$ i.e.,
assume that there are matrices $R,S$ which are functions on
$\Sigma$ such that
$$\Psi([T^A\oplus (dx^{\mu}\otimes T_B)])=
(R_C^A T^C)\oplus (dx^{\mu}\otimes (S_B^DT_D)).$$
We determine conditions which insure that $\Psi$ is a canonical
automorphism of $E.$
According to Lemma $\ref{cov}$ of section 2.1 such an automorphism will be
canonical iff its components satisfy condition $(ii)$ of the Lemma. \\

Observe that $\Psi_A = u_A\circ \Psi=R_A^Cu_C,$
$\Psi^A_\mu = w^A_{\mu}\circ \Psi=S^A_Dw^D_{\mu}$ and

$$\frac {\partial \Psi_A}{\partial u_B}=R^B_A
\quad ,\quad \frac {\partial \Psi^A_\mu}{\partial w^B_{\nu}}=
S^A_B\delta_{\mu}^{\nu}.$$
Consequently, if the matrices satisfy the conditions
$$\overline W_{AB}=W_{CD}R^C_AR^D_B \quad ,\quad
\overline W^{AB}= W^{CD}S^A_CS^B_D$$
where $\overline W_{AB}$ are the components of the tensor $W$ relative to
a new basis $\overline T_A=M_A^CT_C$ and $\overline T^A=(M^{-1})^A_DT^D,$
then
$$\tilde\omega_{AB}=\omega(\tilde \theta_A,\tilde \theta_B)=\overline W_{AB}=
W_{CD}R^C_AR^D_B=\omega(\theta_C,\theta_D)
\frac {\partial \Psi_A}{\partial u_C}
\frac {\partial \Psi_B}{\partial u_D}$$
where we have dropped the volume form $\nu$ from the definition of the
components of $\omega$, as they were defined in section 2.1, for simplicity.
Similarly, we must have
$$\tilde\omega^{A,B}_{\mu,\nu}=\omega(\tilde \theta^A_{\mu},\tilde
\theta^B_{\nu})
=\delta_{\mu\nu}\overline W^{AB}=
W^{CD}\frac {\partial \Psi^A_\mu}{\partial w^C_{\lambda}}
\frac {\partial \Psi^B_{\nu}}{\partial w^D_{\rho}}
\delta_{\lambda \rho}$$
which is the same as
$$\tilde\omega^{A,B}_{\mu,\nu}=
\omega^{C,D}_{\lambda,\rho}
\frac {\partial \Psi^A_{\mu}}{\partial w^C_{\lambda}}
\frac {\partial \Psi^B_{\nu}}{\partial w^D_{\rho}}.$$

Notice that these computations will be consistent if we require that
$R=M$ and $S=M^{-1}$ since $M$ is the matrix transforming the basis
$\{T_A\}$ to $\{\overline T_A\}$, and since we require that $W$ and
$\tilde {W}$ be tensors.
Moreover we must also have that $M$ be orthogonal if we want the transformed
basis to remain $\mu$-orthonormal. These remarks give us the conditions
required in order that a linear automorphism be canonical.

If $G$ is a Lie group and $M:G\rightarrow O(n)$ is a {\em
representation} of $G$ by orthogonal matrices then there is a representation
$\Phi$ of $G$ via canonical automorphisms of $E=V^*\oplus (T^*\Sigma \otimes
V)$ defined by $$\Phi(g)([T^A\oplus (dx^{\mu}\otimes T_B)])= (M(g)_C^A
T^C)\oplus (dx^{\mu}\otimes ([M(g)^{-1}]_B^D T_D)).$$ The fact that
$\Phi:G\rightarrow Aut(E)$ is a {\em group homomorphism} is a consequence of
the fact that $M$ defines a linear left action of $G$ on $V$ via $$g\cdot
T_A=[M(g)^{-1}]^B_A T_B$$ with a corresponding linear left action of $G$ on
$V^*$ defined by $$g\cdot  T^A=M(g)^A_BT^B.$$  The following theorem is a
consequence of these remarks:

\begin{theorem} \label {Poisson}
For each orthogonal $n\times n$ matrix $M$ there is a {\bf canonical} gauge
automorphism  $\Psi_M$ of the bundle
$V^*\oplus (T^*\Sigma \otimes V)\longrightarrow \Sigma$
which is linear on fibers of the bundle and which transforms the basis
$\{T^A\oplus(dx^{\mu}\otimes T_B)\}$ via
$$\Psi_M([T^A\oplus (dx^{\mu}\otimes T_B)])=
(M_C^A T^C)\oplus (dx^{\mu}\otimes ((M^{-1})_B^DT_D)).$$
Moreover, if $M:G\rightarrow O(n)$ is a representation of a Lie group $G$
by orthogonal $n\times n$ matrices, then the mapping
$\Phi:G \rightarrow Aut(V^*\oplus (T^*\Sigma \otimes V))$ defined by
$\Phi(g)=\Psi_{M(g)}$ for $g\in G$, is a representation of $G$ by canonical
automorphisms of $V^*\oplus (T^*\Sigma \otimes V).$ It follows that the
space of local functionals defined on sections of the bundle
$V^*\oplus (T^*\Sigma \otimes V) \rightarrow \Sigma$ admits a reduction
as does the complex $\{\Omega^{k,0}_c[J^{\infty}(V^*\oplus (T^*\Sigma \otimes
V))], k=0,1,2,\cdots,n\}.$
\end{theorem}

\begin{remark} It is not difficult to show that Ikeda's Lagrangian given above
is invariant under the action of the Lie group $G$ defined in Theorem
$\ref{Poisson}$. \end{remark}

\noindent
\textbf{\large Acknowledgements}
The authors would like to thank Jim Stasheff for his useful remarks on a draft
of this paper.

\providecommand{\bysame}{\leavevmode\hbox to3em{\hrulefill}\thinspace}

\end{document}